\documentclass{phbauth}        
\usepackage{graphicx}

\begin{document}

\begin{frontmatter}

\title{Scanning tunneling spectroscopy with superconducting tips of Al}

\author[address1]{I. Guillam\'{o}n, H. Suderow \thanksref{thank1}, S. Vieira}
\author[address2]{and P. Rodi\`{e}re}

\address[address1]{
Laboratorio de Bajas Temperaturas, Departamento de Fisica de la
Materia Condensada, Instituto de Ciencia de Materiales Nicolás
Cabrera, Universidad Aut\'{o}noma de Madrid, E-28049 Madrid, and
Instituto de Ciencia de Materiales de Madrid, Consejo Superior de
Investigaciones Cient\'{i}ficas, Campus de Cantoblanco, E-28049
Madrid.\thanksref{thank1}}

\address[address2]{
Institut N\'{e}el \\ CNRS / UJF, 25, Av. des Martyrs, BP166, 38042 Grenoble Cedex 9, France}

\thanks[thank1]{
* Corresponding author: hermann.suderow@uam.es}

\bigskip

\begin{abstract}
We present Scanning Tunneling Spectroscopy measurements at 0.1 K using tips made of Al. At zero field, the atomic lattice and charge density wave of 2H-NbSe$_2$ are observed, and under magnetic fields the peculiar electronic surface properties of vortices are precisely resolved. The tip density of states is influenced by the local magnetic field of the vortex, providing for a new probe of the magnetic field at nanometric sizes.
\end{abstract}


\end{frontmatter}

\section {Introduction}

The development of new experimental methods for scanning tunneling microscopy and spectroscopy (STM/S) at very low temperatures has opened important paths to obtain fundamental information at the nanoscale. Studies with unprecedented detail can be made thanks to the construction of low temperature STM/S apparatus. For instance, the temperature stability of STS experiments in dilution refrigerators is excellent and allows for long term studies of the electronic properties of surfaces with a resolution in energy which is orders of magnitude below characteristic superconducting energies. Not surprisingly, detailed measurements in different kinds of superconducting materials have unraveled the presence of local size inhomogeneities of the superconducting state \cite{Fischer07,Guillamon07}. These inhomogeneities reveal a number of different effects either of intrinsic nature or due to the presence of defects or impurities. Intrinsic effects are related to multiband properties of the Fermi surface in 2H-NbSe$_2$\cite{Guillamon07}, and to the pseudo gap of the High T$_c$ superconductors\cite{Fischer07}. Impurities or defects produce electron scattering near the surface, whose energy and position dependence permits obtaining important information about local pair breaking effects or the bandstructure of the High T$_c$ superconductors\cite{Yazdani97,Hoffman02}.

One of the most promising future developments is the use of superconducting tips\cite{Stip1,Suderow02,Rodrigo04b,Kohen06}. To see their advantages, let us write down the expression for the tunneling current vs. bias voltage $I(V)$ that is typically used to explain tunneling spectroscopy experiments \cite{Fischer07}. As it is well known,

\begin{eqnarray*}
I(\mathbf{r_0},V) \propto \int dE [f(E-eV)-f(E)] \\
 \times N_T(\mathbf{r_0},E-eV) N_S(\mathbf{r_0},E) T(\mathbf{r_0},E,eV)
\end{eqnarray*}

where $\mathbf{r_0}$ is the position of the tip apex on the sample's surface, $f$ the Fermi function, $N_T$ and $N_S$ the densities of states of tip and sample respectively and $T$ the tunneling matrix element. When the tip is normal the whole position and voltage dependence is usually assumed to be due to the superconducting density of states of the sample $N_S$, and it is easy to find the generally accepted result of a tunneling conductance $G(V)=dI/dV$ that is directly proportional to the convolution between $N_S(\mathbf{r_0},E)$ and the derivative of the Fermi function. When the tip is superconducting, its density of states strongly varies in the energy range of interest. For example, a junction at zero temperature between two isotropic BCS superconductors (note that in STM/S the insulating barrier is vacuum), will have zero current until the bias voltage reaches the sum of gaps of tip $\Delta_T$ and sample $\Delta_S$ (until $V=\Delta_T+\Delta_S$). In the case of having a sample with a density of states differing from BCS theory, it is easy to de-convolute $N_S(E)$ directly from $G(V)$, provided $N_T(E)$ is well known. In addition, the high quasiparticle peaks at $N_T(\Delta_T)$ may in some cases enhance small features of $N_S(E)$, making their measurement easier\cite{Rodrigo04c}. On the other hand, eventual local changes in the superconducting densities of states of the tip $N_T$, produced, for instance, by a varying local magnetic field $b(\mathbf{r_0})$ at the sample's surface, will produce variations in  $N_T(\mathbf{r_0},E)$ that can be used to measure $b(\mathbf{r_0})$. Moreover, at finite temperature, the rounding of the Fermi edge produces additional peaks in $G(V)$ at $V \approx \Delta_T-\Delta_S$, whose height exponentially depends on temperature. Obviously, the Josephson effect, absent in studies made with normal tips, can be readily measured using STM with superconducting tips. Although the experiment is not straightforward, it opens many new possibilities \cite{Rodrigo04,Stip2,Stip3}.

Up to now, tips of Nb and Pb have been used to obtain important new insight into several superconducting materials \cite{Stip1,Suderow02,Rodrigo04b,Kohen06,Crespo07}. Here we present first successful attempts in using superconducting tips of Al.

\section {Experimental}

We use a STM set-up in a dilution refrigerator which allows for scanning ranges up to two microns and has a resolution in energy of 15 $\mu$V\cite{Rodrigo04,Suderow04}. The set-up has been tested previously with several measurements on different superconducting systems. The sample holder is located below the tip and can be moved in-situ by a mechanical pulling mechanism. We glue to the sample holder a sample of the same material as the tip (Al), a sample of Au to test the tunneling characteristics of the tip and a sample of 2H-NbSe$_2$. The magnetic field is always applied parallel to the tip and perpendicular to the sample surface.

\begin{figure}[btp]
 \begin{center}\leavevmode
\includegraphics[width=0.9\linewidth]{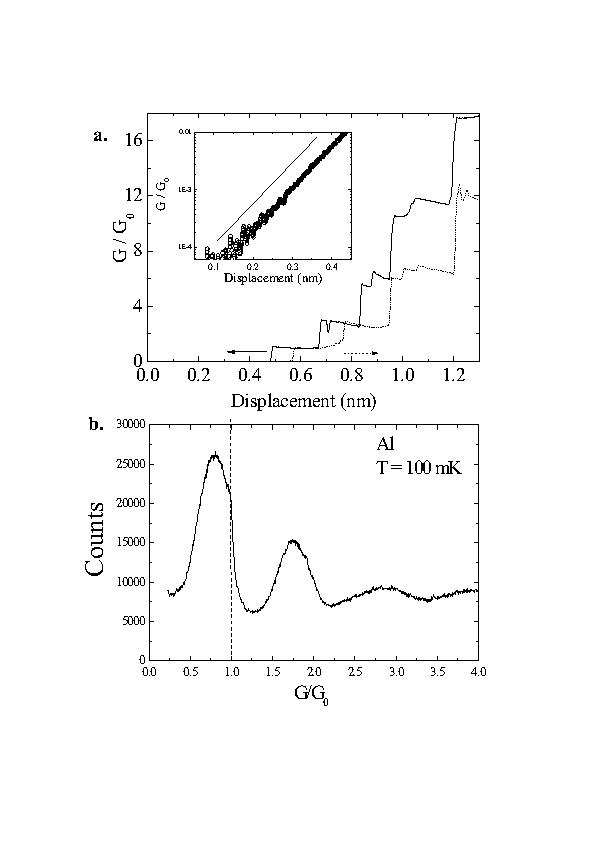}
\vskip -1.5cm
\caption{In a. we show typical conductance vs. distance curves close to the atomic contact regime obtained between tip and sample of Al (line is for indentation and dashed line for retraction of the tip). The conductance is normalized to the quantum of conductance G$_0$=h/2e. In the inset we show the conductance vs. distance obtained in the tunneling limit. In b. we show conductance histograms obtained from thousands of curves similar to the ones shown in a.
 }\label{Fig1}\end{center}\end{figure}

The procedure for obtaining the tips is described in Ref.\cite{Rodrigo04b}. In summary, we position a tip of Al on top of the Al sample, cut the feedback loop, and bring in a controlled way the tip into contact by moving the Z piezo. The surface oxide layer around the tip is easily removed by a repeated indentation procedure, in which the tip is indented up to one micron into the sample. At the same time, additional smaller sized (up to a hundred nm) pushing and pulling ramps are added to the Z piezo signal. In this way, the contacting region is mechanically annealed and shaped\cite{UR97,Agrait95,Suderow02}. During the final pulling process, the formation of atomic size contacts is followed in detail. These contacts lead to a characteristic step like conductance vs. distance $G(z)$ curves, shown in Fig.\ref{Fig1}a. In the tunneling regime, when the contact is fully broken, $G(z)$ has a clear exponential dependence (inset of Fig.\ref{Fig1}a), from which a work function of several eV is extracted. Thousands of $G(z)$ curves are recorded to make histograms that count the values of the conductance found during the experiment. The result is shown in Fig.\ref{Fig1}b, and gives, as found previously \cite{Rev.Nic}, a bell shaped curve with a maximum at about 2/3 of the quantum of conductance $G_0$=h/2e, and a strong decrease at $G_0$. Subsequent smaller peaks are found at values close, but not equal, to multiples of $G_0$. All these features, as well as the form of the curves shown in Fig.\ref{Fig1}a, are related to the atomic orbital properties of the Al atoms \cite{Rev.Nic,Setal97}. Tips with different shapes can be built and moved in-situ to a normal sample for its characterization and then further to the superconducting sample under study. $G(V)$ curves obtained using this procedure on Au have been discussed previously, and are clean, with a negligible density of states close to the Fermi level, and fully explained by isotropic BCS theory using superconducting gaps that range between 0.175 mV and 0.2 mV\cite{Rodrigo04b,Rodrigo04}.

\begin{figure}[btp]
\begin{center}\leavevmode
\includegraphics[width=1\linewidth]{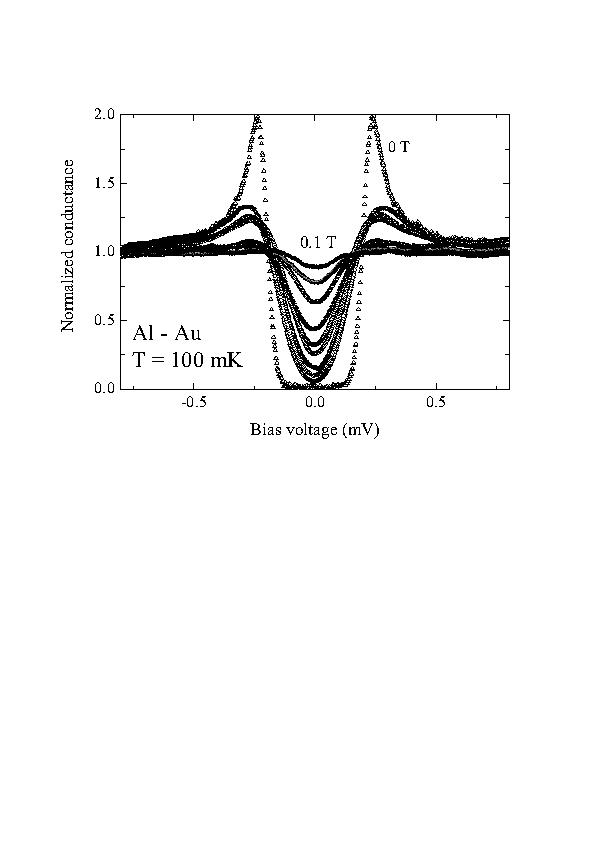}
\vskip -5.2cm
\caption{The magnetic field behavior of the tunneling conductance, normalized to the high bias value, of a tip of Al with a critical field of about 0.1 T. Data have been taken at zero field (triangles), and then from 0.02 T up to 0.1 T, in steps of 0.01 T.}\label{Fig2}\end{center}\end{figure}

Typically, Al tips remain superconducting up to critical fields H$_{c,tip}$ of some hundreds of Gauss. In Fig.\ref{Fig2} we show the tunneling characteristics obtained after having moved in-situ a tip to the normal sample of Au. Superconducting features are observed here up to H$_{c,tip}$ = 0.1 T, although the critical field of Al is of about 0.01 T. The same kind of enhancement of the critical field, has been observed previously in Pb\cite{Rodrigo04b}. Indeed, application of a magnetic field on superconducting tips leads to a curious system in which the superconducting properties strongly vary as a function of the position with respect to the tip apex \cite{Petal98}. Superconductivity is destroyed in the bulk by the applied field, but survives at the apex up to magnetic fields that can be two orders of magnitude higher than the bulk critical field \cite{Suderow02,Rodrigo03}. As the lateral dimensions of the tip close to the apex are easily well below the London penetration depth, superconductivity remains there up to high fields, provided the tip is long and sharp enough so that the proximity effect of the bulk does not destroy superconductivity. H$_{c,tip}$ is found to strongly depend on the geometry of the tip \cite{Rodrigo04b,Suderow02}, and highest values are obtained for the sharpest tips. Ginzburg-Landau calculations have carefully considered the way the order parameter varies when approaching the apex \cite{MFD01,Elmurodov06}.

\section {STM/S in 2H-NbSe$_2$ with superconducting tips of Al.}

\begin{figure}[btp]
 \begin{center}\leavevmode
\vskip -0.5cm
\includegraphics[width=1\linewidth]{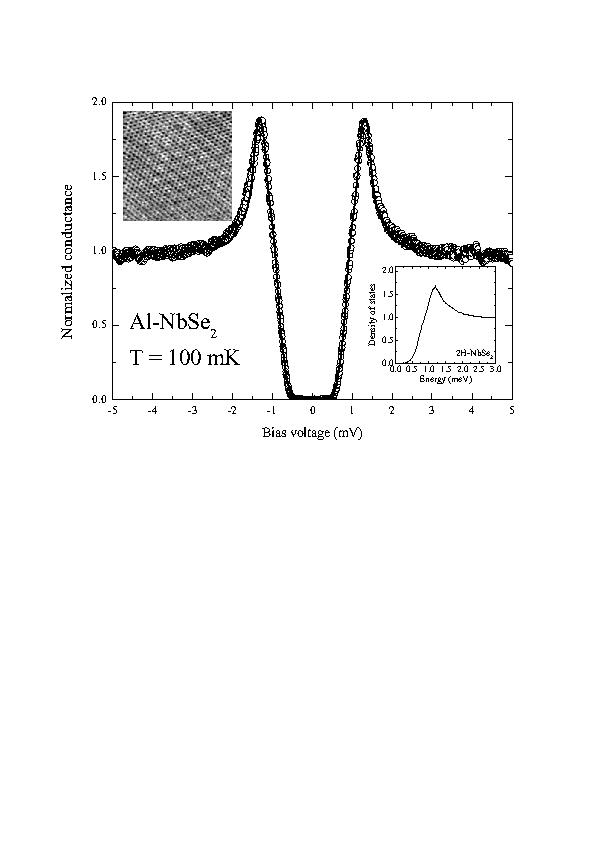}
\vskip -5.2cm
\caption{In the upper left inset we show atomic resolution images made with a tip of Al at zero field. The atomic Se surface lattice of 2H-NbSe$_2$ is clearly viewed, as well as the charge density wave order characteristic of this material. In the figure we show the tunneling $G(V)$ (normalized to its high bias value) obtained in a sample of 2H-NbSe$_2$ with an Al tip at zero magnetic field. Convolution of the density of states shown in the lower right inset, which corresponds to the density of states of 2H-NbSe$_2$ typically measured with normal tips, with a simple BCS density of states with $\Delta$=0.2 mV leads to the line in the figure, which closely follows experiment.}\label{Fig3}\end{center}\end{figure}

In Fig.\ref{Fig3} we show the tunneling characteristics obtained in 2H-NbSe$_2$ at zero field using a tip of Al. The tunneling conductance vs. bias voltage G(V) is at first sight rather similar to the result obtained using normal tips\cite{Hess89,Hess90,Hess91}. A closer look shows however that the quasiparticle peaks are slightly steeper, and located at somewhat higher bias voltages. The density of states of 2H-NbSe$_2$ is very easily de-convoluted from the experimental tunneling conductance data using a BCS expression with a gap of 0.2 mV for the tip (lower right inset of Fig.\ref{Fig3}). When increasing temperature the curves remain qualitatively the same, with some additional broadening. Actually, the features expected in tunnel junctions between two different superconductors at $\Delta_T-\Delta_S$ produced by thermal broadening are not observed in our experiment, because superconductivity in Al is lost already at 1.2 K. This is a significant difference with respect to the results obtained in the same material using Pb tips in Ref.\cite{Rodrigo04c}, where the T$_c$'s of tip and sample are of the same order.

\begin{figure}[btp]
\vskip -0.5cm
 \begin{center}\leavevmode
\includegraphics[width=1.1\linewidth]{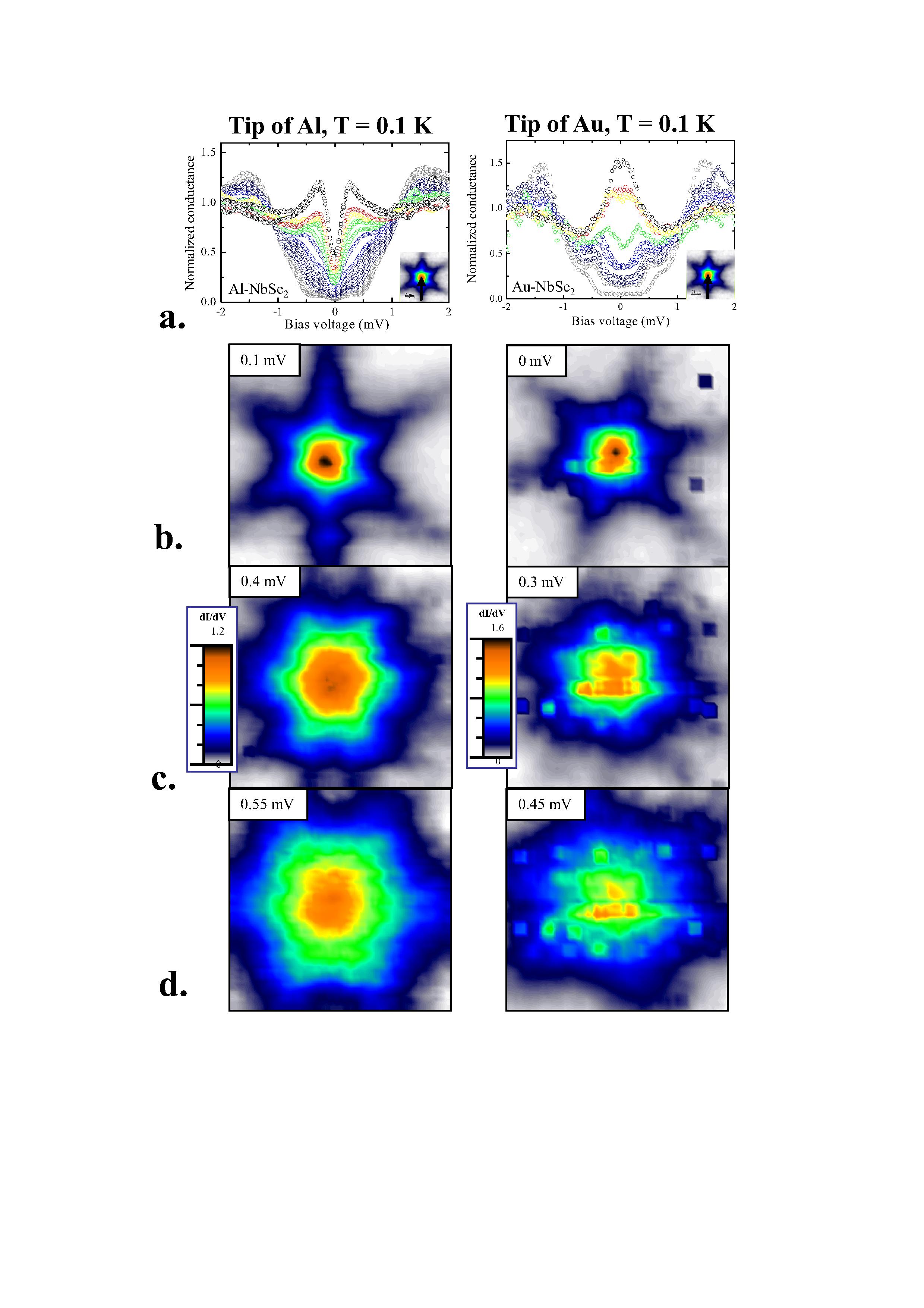}
\vskip -2.5cm
\caption{STS vortex imaging with a tip of Al (left), compared to imaging with a normal tip of Au (right), at 0.1 K. In \textbf{a} we show tunneling conductance curves obtained along a line of about 60 nm that goes into the vortex core, following one of the rays of the star shown in \textbf{b} (top to bottom is from the center to out of the vortex core). Below, we show images of a single vortex at different bias voltages, marked at the upper left corner of each panel. Color scale corresponds to the conductance normalized at high bias. In \textbf{b}, the characteristic star shape of the local density of states of a vortex is visualized. When increasing the bias voltage, the rays of the star split (\textbf{c}) and the star turns by 30$^\circ$ (\textbf{d}). }\label{Fig4}\end{center}\end{figure}

In the mixed state, at a field below H$_{c,tip}$, we clearly observe vortices, as shown in Fig.\ref{Fig4}, where we compare STS results taken at 0.03 T with a superconducting Al tip (left, its H$_{c,tip}$ is of 0.05 T), and results found when using a normal tip (Au, right). The $G(V)$ curves along a line into the vortex core are presented in Fig.\ref{Fig4}a. Let us recall that, as shown previously in Refs.\cite{Hess89,Hess90,Hess91,Hayashi96}, the electronic density of states of a vortex in 2H-NbSe$_2$ is very peculiar, and reflects the localized electronic states that appear within vortex cores as well as the intrinsic anisotropic superconducting properties of this material. A peak develops at the Fermi level density of states at the center of the vortex core that moves to higher voltages when leaving the core (right panel of Fig.\ref{Fig4}a). Additional peaks appear at different voltage values and positions, and the STS images show a strong dependence on bias voltage. At the Fermi level (right panel of Fig.\ref{Fig4}b), a characteristic star is observed, whose rays split into two around 0.3 mV (right panel of Fig.\ref{Fig4}c) and lead to another star shape structure that is turned by about 30$^\circ$ at 0.45 mV (right panel of Fig.\ref{Fig4}d). When the measurement is with a tip of Al, the superconducting density of states of Al is superposed to the local density of states of 2H-NbSe$_2$. Therefore, there is a sharp downturn of the tunneling conductance close to the Fermi level and inside the vortex core (left panel of Fig.\ref{Fig4}a). STS images constructed with the tip of Al show qualitatively the expected features of the local density of states of 2H-NbSe$_2$, namely the star shape, ray splitting and turning of the star (left panels of Fig.\ref{Fig4}b, c and d respectively), at, however, somewhat higher bias voltages due to the gap in the density of states of the tip.

\begin{figure}[btp]
 \begin{center}\leavevmode
\includegraphics[width=0.9\linewidth]{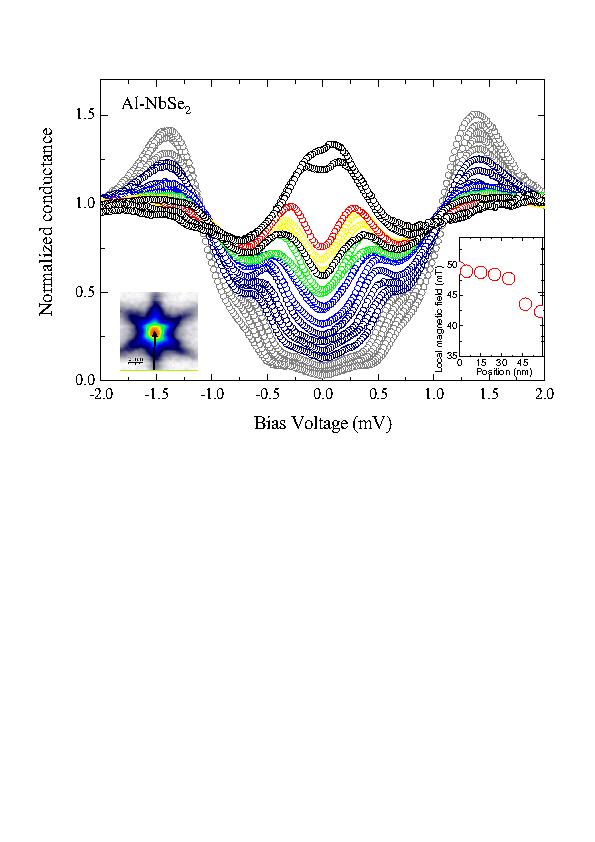}
\vskip -5cm
\caption{Tunneling conductance as a function of the position taken with a tip of Al at 0.04 T. At the vortex core, we find that there is no gap structure corresponding to the tip (top curves of the figure), and that the gap of the tip slowly opens when leaving the vortex core (bottom curves of the figure). The data have been taken along a ray of the star shape vortex, as in previous Fig.\protect\ref{Fig4}a, i.e. along the arrow schematically shown in the bottom left inset. The tip acts as a local probe of the magnetic field, whose position dependence is shown in the right inset.}\label{Fig5}\end{center}\end{figure}

The variation of the magnetic field at the surface $b(\mathbf{r_0})$ in and around vortices (along distances of the order of several tens of nm) at fields of several tens of mT is relatively small, of only around 10\% - 20\%. Detection of the effect of such small changes in $b(\mathbf{r_0})$ on the tunneling conductance of the tip is rather cumbersome, and requires the use of tips whose superconducting density of states $N_T(E)$ strongly depends on the magnetic field. Previous careful measurement of $N_T(E)$ on a normal sample is obviously crucial. In Fig.\ref{Fig5}, we show data taken along a line into a vortex core at 0.04 T with an Al tip whose superconducting density of states shows a stronger field dependence than in the previous experiment. At the center, we observe the zero bias peak expected with a normal tip (top curves of Fig.\ref{Fig5}). However, when leaving the center, the superconducting gap of Al appears around the Fermi level (bottom curves of Fig.\ref{Fig5} are not flat around zero bias, as expected when using a normal tip). The local magnetic field sensed by the tip is then easily computed by de-convoluting the Fermi level density of states $N_T(\mathbf{r_0},E=0)$ of Al from $G(\mathbf{r_0},V)$ (this removes effects due to the density of states of 2H-NbSe$_2$) and using the previously measured dependence of $N_T(E=0)$ as a function of the magnetic field. The result (inset of Fig.\ref{Fig5}) has a large associated absolute uncertainty, which we estimate to be about 50\% in the absolute value of the magnetic field. However, it gives first data of the magnetic field close to the vortex cores, and demonstrates the feasibility of measuring the local magnetic field $b(\mathbf{r_0})$ with a superconducting Al tip as a probe. In addition, the overall order of magnitude of the variation agrees well with the expected variation of the magnetic field using reasonable values for the London penetration depth in this compound \cite{Fletcher07}.

\section {Summary and outlook}

In summary, we have shown STM/S experiments on the surface of 2H-NbSe$_2$, which demonstrate that Al tips are very valuable local probes. The vortex lattice is viewed with great precision, and the superconducting properties of the tip allow for the obtention of both its electronic and magnetic properties. Future developments should also lead to scanning spectroscopy studies of the local Josephson effect. Its observation on Al with STM was indeed already reported in Ref.\cite{Rodrigo03}, although further work is needed to exploit its possibilities.

On more general grounds, it is clear that increasing the palette of elements from which tips can be made is highly desirable. As a matter of fact, it is well known that tunneling conductance critically depends on the tip orbital states \cite{Tersoff83,Tersoff85}. Therefore, different and complementary results should be obtained at the atomic level by using tips with significantly different available orbital states. The use of light elements, as Al, which only has s and p orbitals, as opposed to the more extended use of transition metal elements, could reveal unexpected insight into the properties of the surfaces under study.

\begin{ack}
We specially acknowledge support from NES program, and discussions with V. Crespo and J.G. Rodrigo. This work was also supported by the Spanish MEC (Consolider and FIS-2004-02897 programs), by the Comunidad de Madrid through program "Science and Technology in the Millikelvin", and by ECOM. The Laboratorio de Bajas Temperaturas is associated to the ICMM of the CSIC.
\end{ack}


\bibliography{LastBib_s}

\end{document}